\documentclass{aa}

\usepackage{txfonts}
\usepackage{graphicx}
\usepackage{natbib}

\newcommand \cxc {{\it Chandra}}
\newcommand \xmm {{\it XMM-Newton}}
\begin{document}
\title{Abundance variations and first ionization potential trends during large stellar flares}

\author{Raanan Nordon \inst{1}
\and Ehud Behar \inst{2,1}}

\institute{Department of Physics, Technion, Haifa 32000, Israel; nordon@physics.technion.ac.il
\and Senior NPP Fellow, Code 662, NASA/Goddard Space Flight Center, Greenbelt, MD 20771; behar@physics.technion.ac.il}


\abstract  
{The Solar First Ionization Potential (FIP) effect, where low-FIP (FIP $\la$10~eV) elements are enriched in the corona relative to the photosphere, while high-FIP abundances remain unchanged, has been known for a long while.
High resolution X-ray spectroscopy has revealed that active stellar coronae show an opposite effect, which was labeled the Inverse-FIP (IFIP) effect. The correlation found between coronal activity and the FIP/IFIP bias suggested perhaps that flaring activity is involved in switching from FIP to IFIP. 
}
{This work aims at a more systematic understanding of the FIP trends during stellar flares and complements an earlier study based on \cxc\ alone.} 
{The eight brightest X-ray flares observed with \xmm\ are analyzed and compared with their respective quiescence states.  Together with six previous flares observed with \cxc, this establishes the best currently available sample of flares. We look for abundance variations during the flare and their correlation with FIP. For that purpose, we define a new FIP bias measure.
} 
{A trend is found where coronae that are IFIP biased in quiescence, during flares show a FIP bias with respect to their quiescence composition. This effect is reversed for coronae that are FIP biased in quiescence.
The observed trend is thus consistent with chromospheric evaporation rather than with a FIP mechanism operating during flares.
It also suggests that the quiescent IFIP bias is real and that the large flares are not the direct cause of the IFIP effect in stellar coronae.
} 
{}

\titlerunning{Abundance variations during flares}
\authorrunning{Nordon \& Behar}
\maketitle
\section{Introduction}

\begin{table*}[!t]
\begin{minipage}{\columnwidth}
\renewcommand{\thefootnote}{\arabic{footnote}}
\caption{\label{tab:obs_table} \xmm\ observations used in this work.}
\begin{tabular}{c c c c c c c}
\hline \hline

Obs. ID &  HD & Other Name & Exposure (ks) &Start Time & Type & Distance (pc)$^{H}$  \\

\hline

148790101 & 16157   & CC Eri  & 22.36 & 2003-08-08 09:21:36 & K7Ve/M4$^{S}$  & 11.51 \\
111520101 & 12230   & 47 Cas & 50.9  & 2001-09-11 02:21:19 & F0V+G$^{\natural}$ & 33.56 \\
112880701 & 19356   & Algol    & 45     & 2002-02-12 04:42:18 & B8V+G8III$^{B}$ & 28.46 \\
111530101 & 129333 & EK Dra & 54.9  & 2000-12-30 14:45:20 & G0V$^{H}$ & 33.94 \\
134540401 & 22468 & HR1099 & 26.42& 2001-08-18 03:47:57 & G5IV+K1IV$^{S}$ & 28.97 \\
49350101   & - & Proxima Cen & 67.41 & 2001-08-12 04:16:02 & M5.5V$^{H}$ & 1.295 \\
56030101   & 131156 & $\xi$~Boo & 59 & 2001-01-19 11:14:41 & G8V$^{H}$ & 6.7 \\
111480101 & 62044 & $\sigma$~Gem & 55.8 & 2001-04-06 16:46:36 & K1III$^{H}$ & 37.48 \\

\hline
\end{tabular}
H - HIPPARCOS catalog \citep{Hipparcos} \\
S - \citet{Strassmeier1993} \\
B - \citet{Budding2004} \\
$\natural$ - X-ray active component is not detected directly in the optical. 
It is assumed to be a fast rotating solar analog \citep{Guedel1998, Telleschi2005}\\
\end{minipage}
\end{table*}

The study of stellar coronae was given a significant boost by the launch of \xmm\ and \cxc, that allowed high resolution X-ray spectroscopy.
Observations revealed that the familiar solar First Ionization Potential (FIP) effect, in which the abundance of low FIP elements is enhanced in the corona compared with high FIP elements \citep{Feldman1992}, does not exist in all stellar corona. Some cases showed a clear inverse effect (IFIP), in which the high FIP elements are enriched over the low FIP ones \citep{Brinkman2001}. Some examples show more complex patterns that could indicate more parameters, other than FIP, may have an effect \citep{Huenemoerder2003, Ball2005}.

Further studies indicated a correlation between coronal activity and the FIP effect. \citet{Audard2003} found that highly active RS CVn binaries, as indicated by their effective coronal temperatures, show an IFIP effect while less active (cooler) coronae show either no effect or a solar FIP effect. \citet{Telleschi2005} found a related result in a sample of solar like stars, where abundances change from IFIP to FIP with the age (and decreasing activity) of the star. On the other hand, \citet{Wood2006} compared the abundances of two K type dwarf binaries of similar basic properties (age, spectral type, rotation period, activity level).  They reported different abundance effects ranging from solar-like FIP to none, or weak IFIP effect, inspite of the similar activity levels. 

Since high coronal activity and temperatures is manifested in frequent flares, the correlation between activity and abundances suggest that flares may affect the FIP pattern in some way. However, in analysis of individual stellar flares, mixed results are obtained differing from target  to target: \citet{Guedel1999}, \citet{Audard2001} and \citet{Raassen2003} found an increase in low FIP abundances during flares on the UX Ari, HR 1099 and dwarf binary AT Mic, respectively. In some other cases the variations in abundances were not FIP related \citep{Osten2003, Guedel2004}, or not detected at all \citep{Raassen2007}. \citet{Nordon2007} analyzed six large flares, on different stars, observed with \cxc\ and found an increase in low FIP abundances during five of the flares and no effect in one case.  These results indicate that during flares, if abundance variations were observed, they tended toward the solar-like FIP bias of the flare abundances relative to quiescence, the opposite of what one might expect from the activity-abundance relations reported by \citet{Audard2003} and \citet{Telleschi2005}. Caution should be applied regarding this last statement as the sample in \citet{Nordon2007} is biased to include only the largest flares.

The definitions of flare and quiescence states are themselves somewhat ambiguous. There is growing evidence that the perceived quiescence state is a superposition of many small (micro) flares. From the statistics of large to medium flares, the distribution of the number of flares per unit time as a function of energy released behaves as a power law: $\frac{\mathrm{d}N}{\mathrm{d}E}=C E^{-\alpha}$ with typical values $\alpha \sim 2$ \citep{Hudson1991, Audard2000, Kashyap2002, Guedel2003, Caramazza2007}. 
\citet{Audard1999} used the {\it Extreme Ultraviolet Explorer} (EUVE) to investigate two of the targets also used in this work (47 Cas \& EK Dra) and found a flare statistical distribution with $\alpha=2.2 \pm 0.2$.
If this statistical law holds down to low energies, it means a large number of flares per typical flare cooling time, which in turn would resemble a continuous emission. It is also important to note that $\alpha = 2$ is a critical value above which the total power released by low energy flares depends on (and requires) a low energy cut-off. $\alpha>2$ can potentially explain all quiescent coronal emission as the result of micro-flaring activity.

Stellar photospheric abundances are not well determined and normally, solar-like abundances are assumed. Even in the solar case, photospheric abundances have been revised significantly in the last 20 years from the often used \citet{Angr}. Most of the variation is in the absolute abundance (relative to H) and less in relative abundances of the common heavy elements. The abundance of Ne is notoriously difficult to determine and a significant revision of it has been suggested \citep{Drake2005}. The photospheric abundance of Ne is especially important for the coronal FIP effect in stars, as the two dominant high-FIP elements observed in X-ray are Ne and O. The uncertainties in assumed photospheric abundances are a constant source of doubt of whether the IFIP effect is real \citep{Sanz2004}.

In this work we continue the work from \citet{Nordon2007}, which we will refer to as paper 1. Here, we analyze the spectra of eight flares obtained from the \xmm\ archive. We selected the brightest flares, in terms of number of counts, that allow for accurate line flux measurements. We then compare the thermal and chemical structure between the flaring and quiescence states seeking abundance variation and FIP trends. Together with the \cxc\ flares, we now have a sample of 14 flares, analyzed using similar methods.

\section{Targets and observations}
The \xmm\ public archive was searched for observations of stars of spectral types A to M that include large flares. Light curves were extracted and examined. The criteria for {\it large} is that enough photons were collected during the flare to allow a detailed spectral analysis. 
In this work, we required at least a total of 3000 first order photon counts, between 6 -- 20 $\AA$, in the RGS instruments combined.
Many flares were found, but we retained only those that could provide statistically meaningful measurements. The sample is therefore likely to be biased toward the larger flares occurring perhaps on the more active stars, but these are our current limitations. 

\begin{figure*}[!hp]
\includegraphics[scale=0.9]{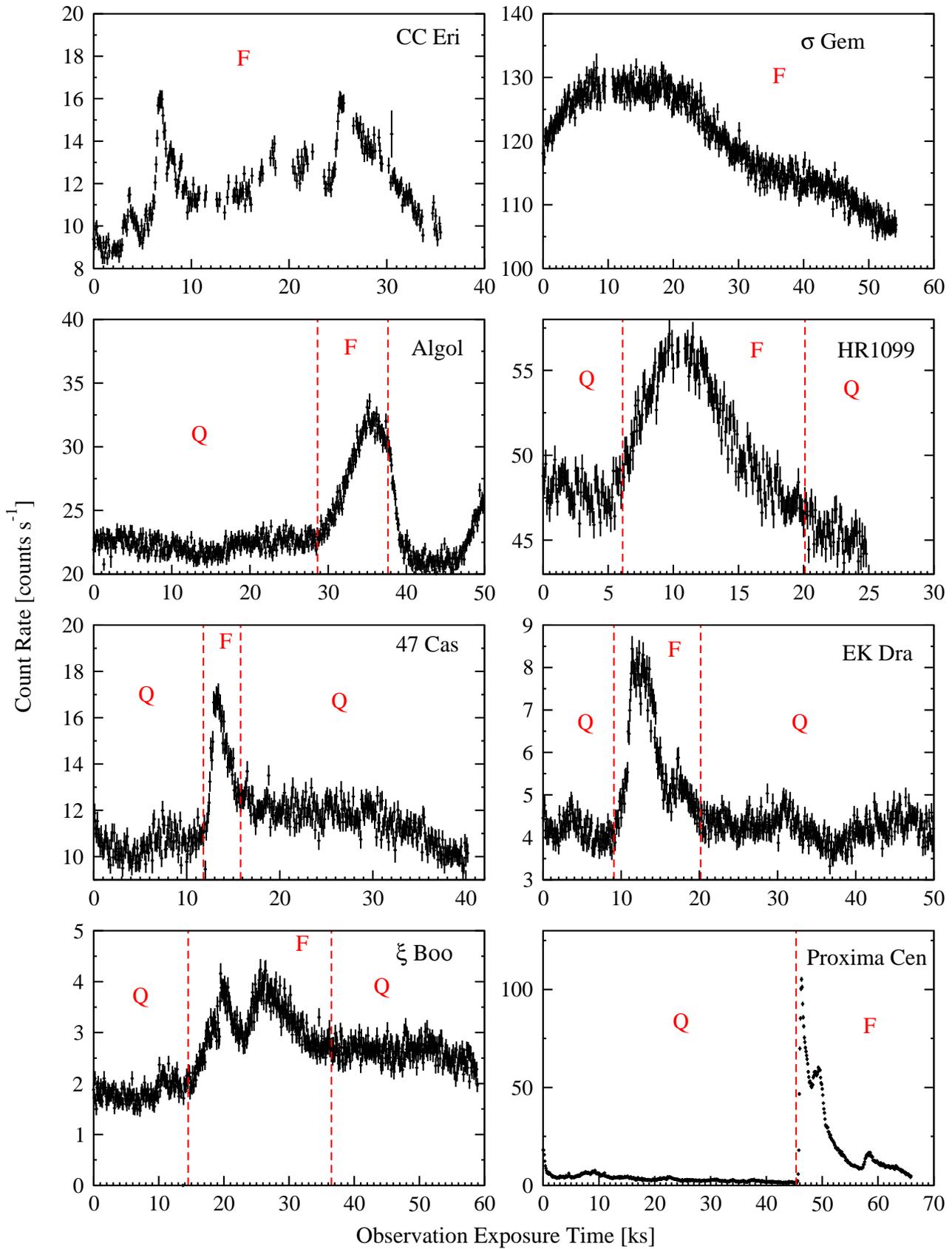}
\caption{Light curves of the flares extracted from EPIC-pn, except for $\xi$~Boo flare that was extracted from EPIC-MOS. Time segments used for flare and quiescence spectra extraction are marked by F and Q respectively. CC Eri and $\sigma$~Gem show no clear quiescence during the observation and therefore a different observation was used as quiescence reference. See text for details.}
\label{fg:lc}
\end{figure*}

The targets and details of the observations are presented in table~\ref{tab:obs_table}.	 Some of these observations were analyzed before. See \citet{Testa2007}, \citet{Crespo2007}, \citet{Schmitt2003}, \citet{Guedel2002a} and \citet{Reale2004} for works relevant to some of the flares analyzed here. 
The light curves of the selected observations are presented in figure~\ref{fg:lc}. The time segments used for flare and quiescence extraction of the spectra are marked on the plots. In the \xmm\ observations of CC~Eri and $\sigma$~Gem, not enough photons were detected in quiescence and quiescence spectra from \cxc\ HETGS observations were used; For CC Eri, the quiescence data from paper 1, and for $\sigma$~Gem a 120 ks archival observation (Obs. ID 5422, 6282). 

\section{Analysis methods}
The present analysis methods are similar to those used in paper 1.  We briefly summarize them and emphasize the differences due to the use of \xmm\ instruments instead of \cxc. The goal is to reconstruct the thermal structure of the plasma and measure the abundances during flaring and quiescence states. The method relies on line flux analysis, which is much more sensitive to temperatures and abundances than the continuum.

In order to investigate FIP biased abundance variations, we are interested in relative, more than absolute, abundances. Therefore, the continuum which is dominated by H and which is typically used to measure the absolute abundances, is of lesser importance. Instead, we measure the abundances relative to Fe, which emits lines from a wide range of plasma temperatures.
Due to the high resolution of the grating instruments, the lines have a very high equivalent width and local continuum uncertainties have a relatively small effect on the measured fluxes. Line fluxes measured from the CCD instruments (lower resolution) are prone to larger systematic errors caused by uncertainties in the continuum determination and increased difficulty in resolving line blends. This should be kept in mind when reviewing the results based solely on the latter.

We define the Emission Measure Distribution ($EMD$) as:
\begin{equation}
EMD(T) = n_e n_H dV/dT
\label{eq:EMD_definition}
\end{equation} 

\noindent where $n_e n_H$ are the averaged $e^-$ and H number densities in the plasma at the temperature interval [T, T+dT]. Since Hydrogen does not emit lines at coronal temperatures, its emission contribution is confined to the continuum. We are interested in abundance and thermal structure {\it variations} and so the more useful quantity for our purpose is the Iron emission measure distribution ($FeEMD$):
\begin{equation}
FeEMD  \equiv n_e n_{Fe} dV/dT=A_{Fe}EMD
\label{eq:FeEMD_definition}
\end{equation}

\noindent where $A_{Fe}$ is the iron abundance relative to H.

\xmm\ comprises several different instruments.  The longer wavelengths (6-38~$\AA$) are covered by the two high resolution reflection grating spectrometers (RGS) that resolve the emission lines. The high energy end up to 15~keV is covered by the two EPIC-MOS and single EPIC-pn CCDs, that offer lower spectral resolution. Since the present targets are all X-ray bright, point sources, EPIC-MOS was often piled-up. This was a much lesser problem in EPIC-pn and so we use only this instrument for the high energies, except in the case of $\xi$~Boo where EPIC-pn data were not available.
This reduces the accuracy of the line fluxes measured at wavelengths shorter than 6~$\AA$.  The K-shell lines of S, Ar and Ca can thus be measured only at CCD resolution. The Si-K lines are just at the end of the RGS band and in a few cases the signal was good enough to allow high-resolution constraints on these lines. 

The observed spectrum is fitted by sets of complete individual-ion spectra simultaneously \citep{Behar2001}, together with a bremsstrahlung continuum, composed of several discrete-temperature components as explained in paper 1. We emphasize again that the fitted continuum serves here only the line flux measurements and is not used in the $EMD$ nor in the abundance analysis.
The extracted line fluxes are listed in tables~\ref{tab:CC_Eri_lf}-\ref{tab:Xi_Boo_lf} in the appendix.

Once a list of line fluxes is compiled, we use the same methods described in paper 1 to measure the $FeEMD$ and the abundances.  In short, 
the set of equations to be solved, for the measured line flux $F^{Zq}_{j i}$ from ion $q$, of element $Z$, due to the transition $j \rightarrow i$ is:
\begin{equation}
  F^{Zq}_{j i} = \frac{A_Z/A_{Fe}}{4 \pi d^2} \int_{T_{0}}^{T_{max}}{P^{Zq}_{ji}(T) f_{Zq}(T) FeEMD(T) \mathrm{d}T }
\label{eq:solve_eq}
\end{equation}

\noindent where $d$ is the distance to the target, $P^{Zq}_{ji}$ is the line emission rate coefficient (per unit electron density), $f_{Zq}(T)$ is the ionic fraction and $A_Z/A_{Fe}$ is the abundance relative to Fe. Rate coefficients are calculated using HULLAC \citep{HULLAC} and ionic fractions for: Fe, Ar, S, Si, Mg are from \citet{Gu2003} while for the other elements \citet{Mazzotta1998} is used.
A parametrized $FeEMD$ is fitted to reproduce the measured flux of a selected set of strong and weakly blended Fe lines. For non-Fe elements, we use pairs of lines from different ionization degrees, in which case their flux {\it ratio} is fitted for. 
In practice, with the RGS, the lowest temperature constraint is obtained from the ratio of the two K-shell charge states of oxygen. 
The emission from O VII drops significantly below $kT=$0.1~keV, giving us no constraints on the $EMD$ at lower temperatures. Therefore 0.1 keV was chosen as the lower limit for the $EMD$ analysis. Once the $FeEMD$ is determined, it is used to calculate the abundances $A_Z/A_{Fe}$ by plugging it back into eq.~\ref{eq:solve_eq}. Elements for which only one ionization degree is detected in a specific observation (i.e. one of the lines is consistent with zero, see line flux tables in the supplement) are not used for $FeEMD$ constraints, but are used for measuring abundances. 

The $FeEMD$ is described by  a stair-case function where $FeEMD$ in a bin represents the averaged $FeEMD$ over the bin temperature range. The calculated errors on the $FeEMD$ include also uncertainties caused by cross-talk between neighboring bins. Increasing the temperature resolution by using narrower $FeEMD$ bins results in larger uncertainties in individual bins. For this reason, we also calculate the useful quantity of the Fe integrated emission measure:

\begin{equation}
FeIEM(T) = \int_{T_0}^{T}{FeEMD(T)\mathrm{d}T}
\label{eq:FeIEM_definition}
\end{equation}

\noindent Integrating the $FeEMD$ over $T$, while accounting for the correlations between bins reduces the uncertainties and provides a clearer picture of the temperatures of excess flare emission, as demonstrated in paper 1 and in the following. To check that the results depend only weakly on the binning and on the functional form of the $FeEMD$, similar to what was done in paper 1, we also fit an $FeEMD$ parametrized as an exponent of a polynomial: $\exp(P(T))$. Evidently, the $FeIEM$ and the abundances calculated using this solution agree extremely well with those obtained from the staircase $FeEMD$. An important advantage of the latter is that it enables the computation of local $FeEMD$ uncertainties, not available with globally parametrized functions.

\section{Results}
\subsection{FeEMD and FeIEM}

\begin{figure*}[!hp]
\includegraphics[scale=0.91]{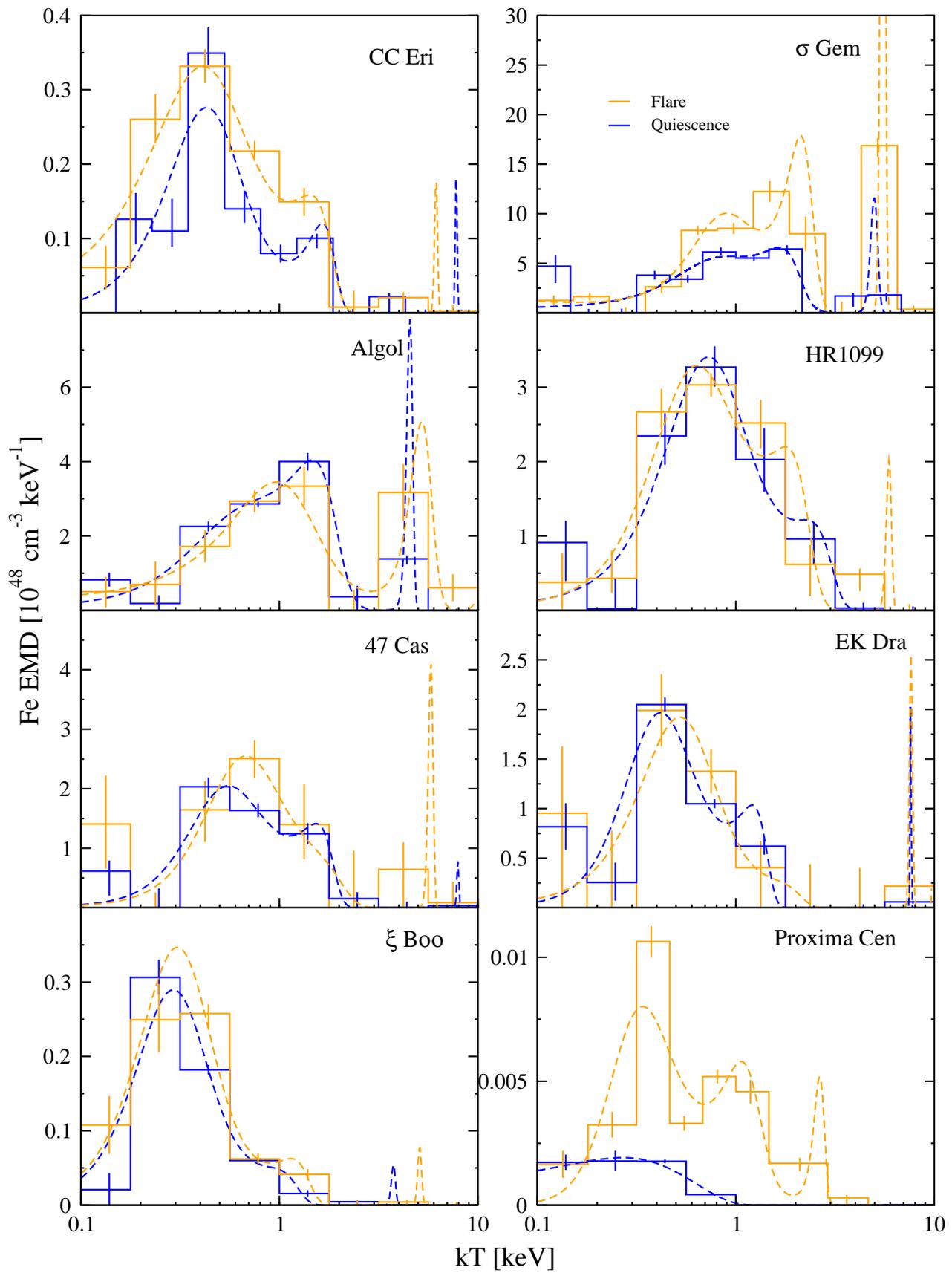} 
\caption{$FeEMD$ solutions for the flare (orange) and quiescence (blue). 
Errors are 1 $\sigma$ and include uncertainties due to correlation between the bins.
Dashed lines represent the polynomial parameterized solutions.}
\label{fg:FeEMD}
\end{figure*}

\begin{figure*}[!hp]
\includegraphics[scale=0.91]{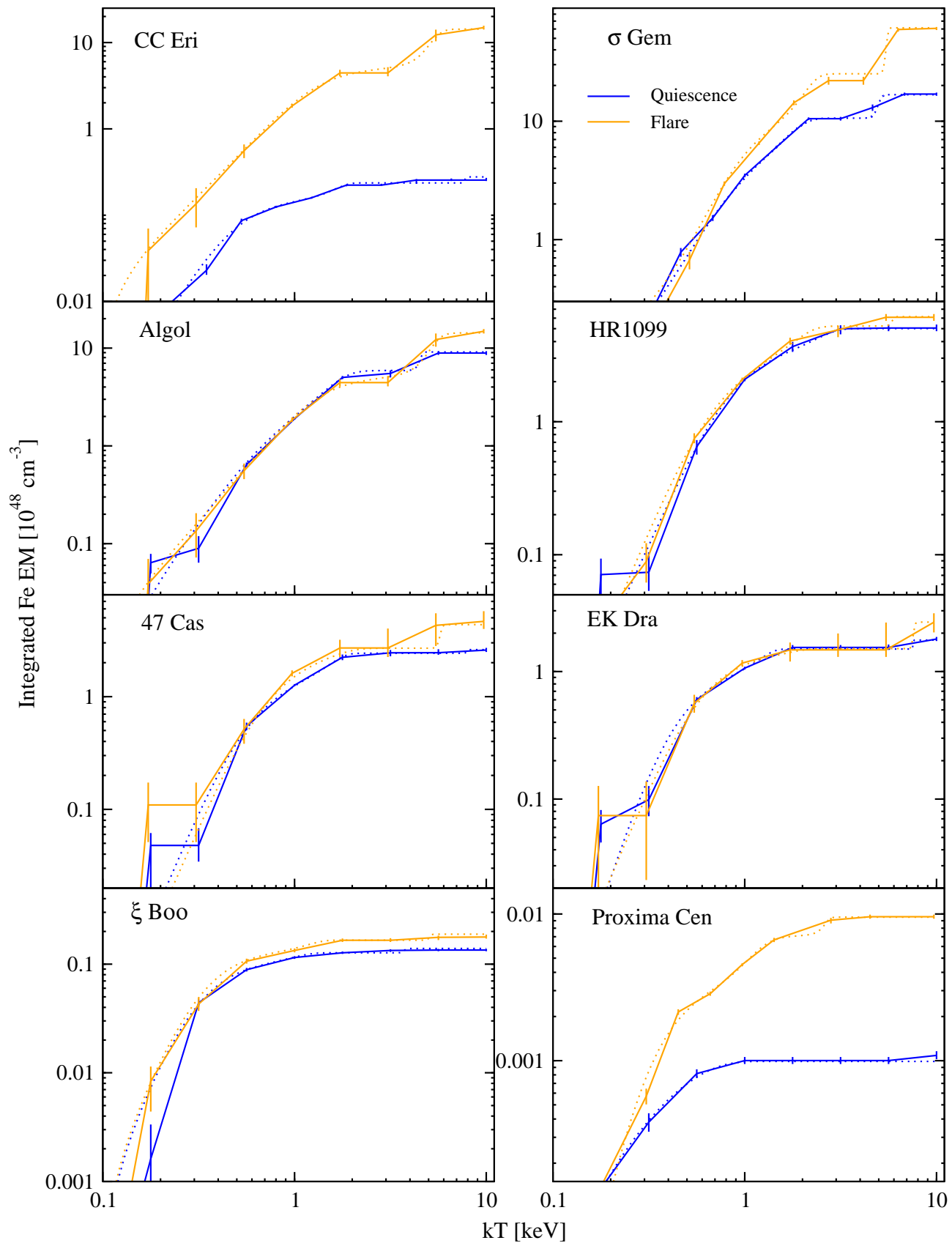}
\caption{$FeIEM$ results for flare (orange) and quiescent (blue) states, namely the progressive integral over the $FeEMD$ of figure \ref{fg:FeEMD}. Plotted errors account for uncertainties due to correlations between bins during integration.
Dotted lines show the results of the integration over the polynomial parameterized $FeEMD$ (see figure~\ref{fg:FeEMD}).}
\label{fg:FeIEM}
\end{figure*}

												
\begin{table*}[!ht]												
\caption{\label{tab:abundances} Abundances relative to Fe												
during flare, quiescence and the flare to quiescence ratio. Quiescence abundances for CC Eri and $\sigma$~Gem are from \citet{Nordon2007}.}
\begin{center}
\begin{tabular}{c | c c c c c c | c c c c c c}												
\hline \hline												
 & \multicolumn{6}{c}{{\bf CC Eri}} & \multicolumn{6}{|c}{{\bf $\sigma$~Gem}} \\												
 & \multicolumn{2}{c}{Quies.} & \multicolumn{2}{c}{Flare} & \multicolumn{2}{c}{Flare/Quies.}										
&\multicolumn{2}{|c}{Quies.} & \multicolumn{2}{c}{Flare} & \multicolumn{2}{c}{Flare/Quies.} \\										
El. & X/Fe &Error & X/Fe & Error & Ratio & Error 
	 & X/Fe &Error & X/Fe & Error & Ratio & Error \\
\hline													
												
C	& -	& -	&47.9	&2.3	& -	& -	& -	& -	&32.1	&2.6	& -	& - \\
N	& -	& -	&16.6	&0.8	& -	& -	& -	& -	&21.3	&1.4	& -	& -\\
O	&72.3	&3.3	&48.6	&1.4	&0.67	&0.04	&44.6	&2.7	&27.5	&1.5	&0.62	&0.05\\
Ne	&21.1	&0.5	&19.5	&0.6	&0.92	&0.04	&20.1	&0.9	&14.0	&0.8	&0.70	&0.05\\
Mg	&1.49	&0.07	&1.77	&0.14	&1.2	&0.1	&1.89	&0.08	&2.2	&0.1	&1.17	&0.09\\
Si	&2.17	&0.09	&2.2	&0.1	&1.01	&0.07	&1.39	&0.06	&1.53	&0.08	&1.11	&0.08\\
S	&1.26	&0.15	&1.03	&0.08	&0.8	&0.1	&0.63	&0.04	&0.81	&0.05	&1.3	&0.1\\
Ar	&0.4	&0.1	&0.11	&0.05	&0.27	&0.16	&0.50	&0.04	&0.23	&0.02	&0.46	&0.05\\
Ca	&0.08	&0.1	&0.00	&0.06	& -	& -	&0.19	&0.03	&0.20	&0.02	&1.0	&0.2\\
Ni	&0.07	&0.01	&0.08	&0.02	&1.0	&0.3	&0.07	&0.01	&0.07	&0.01	&0.9	&0.2\\
												

\hline 
 & \multicolumn{6}{c}{{\bf Algol}} & \multicolumn{6}{|c}{{\bf HR1099}} \\												
\hline													
C	&1.4	&1.0	& -	& -	&-	&-	&24.3	&2.5	&31.7	&2.4	&1.3	&0.2\\
N	&20	&1	&24	&3	&1.2	&0.2	&6.3	&0.8	&7.9	&0.6	&1.3	&0.2\\
O	&21.6	&0.7	&22.5	&1.6	&1.04	&0.08	&26.7	&2.2	&33	&1	&1.2	&0.1\\
Ne	&9.7	&0.5	&10.6	&0.9	&1.1	&0.1	&7.0	&0.7	&9.0	&0.5	&1.3	&0.1\\
Mg	&1.7	&0.1	&1.7	&0.3	&1.0	&0.2	&2.4	&0.4	&1.8	&0.2	&0.8	&0.1\\
Si	&1.26	&0.15	&0.9	&0.3	&0.7	&0.3	&1.7	&0.3	&1.52	&0.08	&0.9	&0.2\\
S	&0.57	&0.06	&0.32	&0.12	&0.6	&0.2	&0.1	&0.3	&0.55	&0.05	&5	&12\\
Ar	&0.33	&0.06	&-	&-	&-	&-	&-	&-	&-	&-	&-	&-\\
Ca	&0.10	&0.04	&0.18	&0.06	&2	&1	&-	&-	&-	&-	&-	&-\\
Ni	&-	&-	&0.07	&0.04	&-	&-	&-	&-	&-	&-	&-	&-\\

\hline 
 & \multicolumn{6}{c}{{\bf 47 Cas}} & \multicolumn{6}{|c}{{\bf EK Dra}} \\												
\hline														
C	&19	&3	&12.0	&5.6	&0.6	&0.3	&9.5	&1.4	&13	&6	&1.3	&0.7\\
N	&7.0	&0.8	&9.4	&2.9	&1.3	&0.4	&3.5	&0.5	&1.9	&1.4	&0.6	&0.4\\
O	&29	&1	&27	&4	&0.9	&0.1	&15.8	&0.9	&18.5	&2.4	&1.2	&0.2\\
Ne	&9.7	&0.6	&8.7	&1.8	&0.9	&0.2	&5.8	&0.4	&8.6	&1.5	&1.5	&0.3\\
Mg	&2.1	&0.2	&2.7	&0.6	&1.3	&0.3	&1.6	&0.2	&2.0	&0.5	&1.3	&0.3\\
Si	&1.00	&0.09	&1.0	&0.3	&1.0	&0.3	&1.0	&0.1	&2.1	&0.9	&2.1	&0.9\\
S	&0.43	&0.06	&0.5	&0.2	&1.1	&0.5	&0.30	&0.06	&0.6	&0.3	&2.1	&1.2\\
Ar	&0.13	&0.05	&0.00	&0.16	&-	&-	&-	&-	&-	&-	&-	&-\\
Ca	& -	& -	&0.1	&0.2	& -	& -	&-	&-	&0.8	&0.5	&-	&-\\
Ni	&0.3	&4.7	&0.05	&0.05	&0.2	&3	& -	&-	&0.16	&0.05	&-	&-\\

\hline 
 & \multicolumn{6}{c}{{\bf $\xi$ Boo}} & \multicolumn{6}{|c}{\bf{Proxima Cen}} \\												
\hline													
C	&7.6	&0.6	&6.6	&0.8	&0.9	&0.1	&24	&3	&32	&2	&1.3	&0.2\\
N	&1.4	&0.2	&1.3	&0.2	&0.9	&0.2	&6.3	&0.8	&7.9	&0.6	&1.3	&0.2\\
O	&11.5	&0.5	&11.3	&0.7	&0.98	&0.07	&27	&2	&33	&1	&1.2	&0.1\\
Ne	&3.6	&0.2	&3.5	&0.3	&1.0	&0.1	&7.0	&0.7	&9.0	&0.5	&1.3	&0.1\\
Mg	&1.6	&0.1	&1.2	&0.1	&0.74	&0.09	&2.4	&0.4	&1.8	&0.1	&0.8	&0.1\\
Si	&1.2	&0.2	&1.0	&0.2	&0.8	&0.2	&1.7	&0.3	&1.52	&0.08	&0.90	&0.15\\
S	&0.41	&0.07	&0.7	&0.1	&1.6	&0.4	&0.1	&0.3	&0.55	&0.05	&5	&12\\
Ar	&0.18	&0.06	&0.11	&0.08	&0.6	&0.5	&-	&-	&-	&-	&-	&-\\
Ca	&0.25	&0.09	&-	&-	&-	&-	&-	&-	&-	&-	&-	&-\\
Ni	&-	&-	&-	&-	&-	&-	&-	&-	&-	&-	&-	&-\\
\hline												
\end{tabular}
\end{center}												
\end{table*}

Figure~\ref{fg:FeEMD} shows the fitted $FeEMD$ for each target, in flare and in quiescence states. The errors on the binned $FeEMD$ are 1-$\sigma$ and include uncertainties due to both line fluxes and correlations between bins. The dashed lines are the fitted $FeEMD$ parametrized as an exponent of a polynomial. The most notable flare $FeEMD$ excess is observed for CC~Eri, $\sigma$~Gem, and Proxima Cen. The other flares become more obvious only upon integration ($FeIEM$).
The polynomial $FeEMD$s give somewhat similar patterns to the bin-form $FeEMD$s.
At high temperatures, typically 5--8~keV, the polynomial model tends to fit a spike, 
since the constraints in this temperature region are dominated by the Fe XXIV and Fe XXV ratio and a single-$T$ component is enough to fix it. Fe~XXVI is just beginning to form and has a weak signal in most flares. Line emissivities of other elements are very low in this temperature range, as they are mostly in bare ion form.

Figure~\ref{fg:FeIEM} shows the cumulative distribution (Iron integrated emission measure, $FeIEM$) from 0.1--10~keV.  Errors are calculated including correlations between bins and thus largely cancel out with integration. This considerably reduces the relative errors and thus provides a useful physical comparison between flare and non-flare. The dashed lines are the $FeIEM$s for the polynomial parametrized $FeEMD$, seen to match well with the binned $FeEMD$ results, within the errors. In poorly constrained $FeEMD$ temperature regions, the solid and dashed curves may slightly diverge, only to re-converge once integrated over the less certain range. 
This is clearly demonstrated in the high-$T$ range, where the polynomial model has a distinct spike, but results in the same $FeIEM$ after integration. The two models always agree particularly well on the total EM, represented by the last point in the $FeIEM$ plots of figure~\ref{fg:FeIEM}. 
This demonstrates that the errors in the $FeEMD$ are clearly dominated by the degeneracy of the $FeEMD$ solution and that the chosen parametrization has very little effect.

From figures~\ref{fg:FeEMD} and \ref{fg:FeIEM}, we see that most of the excess emission during the flare originates from temperatures above $kT$ = 1~keV. In most cases, the time averaged emission measure, under 1~keV, during the flare, is not significantly different from the quiescence emission measure. Only CC Eri, $\sigma$~Gem and Prox Cen have significant EM excess up to 1~keV as is evident in figure~\ref{fg:FeIEM}. If we think of the flare in terms of impulsive  heating and gradual radiative/conductive cooling, then the time averaged $EMD$ excess is due to plasma cooling through a given temperature. Flares that exhibit high total EM excess, but little low-$T$ excess, raise the intriguing question of how is the EM lost, without leaving a trace of cooling.
A low $EMD$ in a given temperature range, as observed in the time averaged data, could result from very rapid cooling through that temperature range. A significant increase in conductive cooling at low temperatures could potentially provide this rapid cooling mechanism, although a special geometry would need to be invoked to make conduction at low-$T$ more efficient than at high-$T$.
Especially in the cases of the flares on HR1099, 47 Cas, EK Dra and $\xi$~Boo, EM loss other than rapid cooling through X-ray sensitive temperatures may be required. Possibilities include rotation of the flare region beyond the limb, or a rapid decrease of density due to expansion. 
In the case of Algol, this could be due to the eclipse hiding the decay phase (see figure~\ref{fg:lc}).

\subsection{Abundances}

\begin{figure*}[!hp]
\includegraphics[scale=0.90]{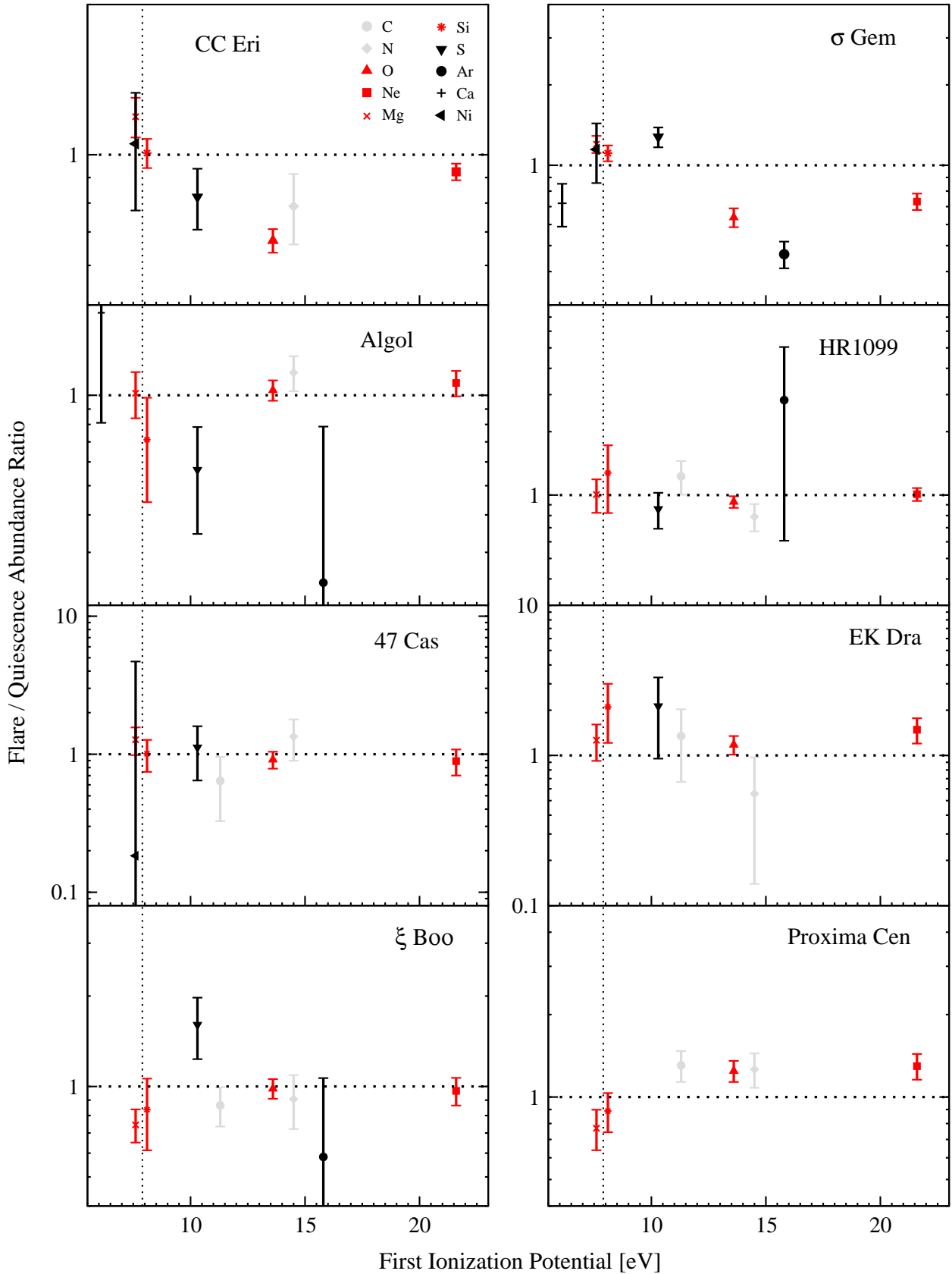}
\caption{Flare abundances relative to quiescence, as a function of FIP. Elements observed with the RGS (O, Ne, Mg, Si) are considered the most reliable and are marked in red. Those observed only with the low-resolution EPIC instruments are marked black. Elements whose abundance determination depends on the $EMD$ below temperatures of $kT=0.1$~keV, for which we did not solve, may suffer from systematic errors and are plotted in gray. The vertical dotted line mark the FIP of of the reference element Fe.}
\label{fg:FIP}
\end{figure*}

The measured abundances relative to Fe, obtained from the binned $FeEMD$, are detailed in table~\ref{tab:abundances}. They are measured relative to Fe using emission lines only and not relative to Hydrogen. 
Note that the abundances are the actual abundances and not relative to solar. The errors on the abundances include both statistical errors on the measured line fluxes and errors due to $FeEMD$ solution uncertainties. One must keep in mind that some of the lines are measured using the EPIC instruments which may add systematic errors. This is especially true for the abundances of S, Ar, and Ca. 
Another source of systematic errors is our inability to resolve the $EMD$ below $\sim$0.1~keV. The lines of N and C, which form largely at these low temperatures, might thus suffer from additional uncertainties. We therefore concentrate on the more reliable abundances of O, Ne, Mg, Si and Fe as reference. Note that abundances obtained from bright X-ray emission lines are constrained even better than the solar photospheric abundance measurements. The abundances obtained from the polynomial $FeEMD$ are identical to those obtained from the binned $FeEMD$, largely within a fraction of a standard deviation. Due to the similarity of the results, the tables only specify the abundances obtained from the binned parametrization, for which we can include $FeEMD$ induced uncertainties.

Figure~\ref{fg:FIP} shows the abundance ratios of flare and quiescence states as a function of FIP. These ratios are derived directly from the X-ray data and require absolutely no knowledge of photospheric abundances. Consequently, they provide a clean measure of abundance variations during and due to the flare.
It should be emphasized that what are defined as `flare' abundances are an $EMD$ weighted average of the flaring region and the background quiescence abundances. 
In fig.~\ref{fg:FIP}, the abundances based on RGS data are considered the most reliable and are highlighted in red. Abundances determined from EPIC data are less certain and are marked in black, as is that of Ni which is based on weak and somewhat blended lines. The abundances of C and N are plotted in gray due to possible emission from unresolved low temperatures that could hamper their measurement. 
As seen in fig.~\ref{fg:FIP}, flares on different stars show different FIP trends ranging from solar FIP to inverse FIP and including no FIP effect at all. Although at first sight this may seem like a totally arbitrary abundance behavior during flares, in the following section we argue that in fact a clear FIP-bias pattern emerges from these plots.

\section{Discussion}
\subsection{The FIP bias}
We can differentiate between three cases: CC Eri and $\sigma$~Gem show a change in the abundances during the flare that resembles the solar FIP effect - low FIP elements are enriched relative to the high FIP elements. These results are similar to what was found in paper 1 for five out of the six flares observed with \cxc, including a different flare on CC Eri. A similar result for $\sigma$~Gem for the same flare, but using a different quiescence reference
was reported by \citet{Nordon2006}. The flare on Proxima Cen, on the other hand shows an inverse-FIP like effect, where it is the high FIP elements which are enriched relative to the low FIP ones. This result confirms the pattern reported by \citet{Guedel2004}. The other five flares show no clear pattern and are consistent within the errors with no abundance variation.

In order to better quantify the abundance variations, a measure for the FIP bias is required. Let us define a FIP bias measure in the following way:

\begin{equation}
	FB = \left< \log A_{Z \mathrm{rel}}(\mathrm{low}) \right> - \left< \log A_{Z \mathrm{rel}}(\mathrm{high}) \right>
\label{eq:FB}
\end{equation}

\noindent Where $\left< \log A_{Z \mathrm {rel}}(\mathrm{low / high}) \right> $  is the mean abundance (in log) of the low/high FIP elements relative to a set of reference abundances. The averaging of the log abundances is a gaussian mean and uncertainties are propagated accordingly.
We use the traditional solar distinction between low (FIP$<$10~eV) and high (FIP$>$10~eV) FIP elements.
A positive FB value indicates a solar-like FIP bias, while a negative FB value indicates an IFIP bias.
The FB defined above can be applied to flare abundances with reference to quiescence ($FB_{FQ}$), or to quiescence abundances with reference to photospheric (e.g., solar; $FB_{QS}$).  In the present analysis, in order to minimize systematic errors in the calculation of the FB, we use only O and Ne as the high-FIP elements and Si, Mg and Fe (implicitly) as the low-FIP elements, thus leaving out the other, less well-constrained abundances.

\begin{figure}[t]
\includegraphics[width=0.48\textwidth]{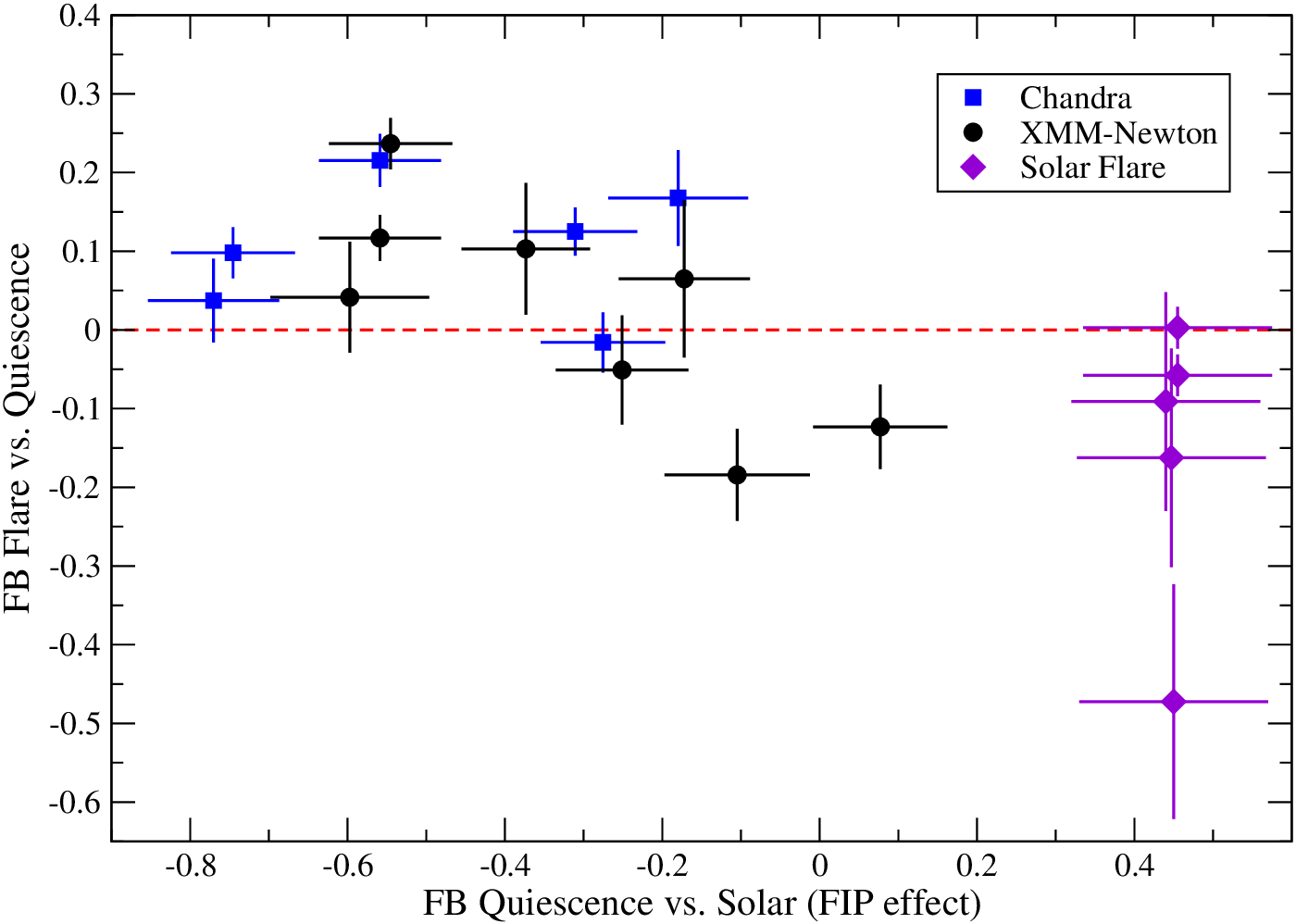}
\caption{The FIP bias measure FB, defined in eq.~\ref{eq:FB}, of the flare plasma relative to quiescence as a function of the FB of the corona in quiescence, relative to Solar (photospheric). Black circles are from \xmm\ observations, blue squares are from \cxc\ (paper 1) violet diamonds are for solar flares.\bigskip \bigskip}
\label{fg:FB_FQ_QS}
\end{figure}

We calculate for each flare $FB_{FQ}$ with respect to its quiescence abundances and also $FB_{QS}$ of the quiescent abundances with respect to the solar photospheric composition, adopting the solar abundances of \citet{Asplund2005}.
Figure~\ref{fg:FB_FQ_QS} shows a plot of $FB_{FQ}$ as a function of $FB_{QS}$ for all \xmm\ flares analyzed in this paper and also for the six flares from paper 1 observed with \cxc. Most stellar flares in the sample tend to exhibit a positive $FB_{FQ}$. In other words, low-FIP abundances increase relative to high-FIP abundances during the flare. Only two flares show a statistically significant negative $FB_{FQ}$: Proxima~Cen and $\xi$~Boo. Five flares are consistent with $FB_{FQ}=0$ to within a standard deviation. The errors plotted in figure~\ref{fg:FB_FQ_QS} reflect the statistical uncertainties of the line fluxes and include also the $EMD$ induced uncertainties.
It is important to emphasize that the calculation of $FB_{FQ}$ suffers from very little systematic errors as it does not require any knowledge of the photospheric abundances. Furthermore, it results from a systematic abundance study from the present work and from paper 1 that uses the same methods and the same atomic data for all flare and quiescence abundances. 

The calculated $FB_{QS}$ values (the x axis in figure~\ref{fg:FB_FQ_QS}) do depend on the set of photospheric abundances selected. We adopt the solar abundances of \citet{Asplund2005} as well as their errors, which are included in (and can dominate) the $FB_{QS}$ errors plotted in figure~\ref{fg:FB_FQ_QS}.  For the sake of discussion, we assume that the stellar photospheres are similar in composition to the Sun. Deviation of the actual photospheric abundances from solar would affect the horizontal position of the data points in fig.~\ref{fg:FB_FQ_QS}, but not their vertical position. Not surprisingly, the flare targets tend to be of the $FB_{QS}<0$ type, i.e.,  IFIP coronae, since large flares are more commonly observed on active coronae, which are IFIP biased \citep{Brinkman2001, Audard2003, Telleschi2005}. 
Note that the two flares observed to have $FB_{FQ}<0$ are also those that erupt on the stars with the highest $FB_{QS}$ values. 

Since X-ray observations of coronal flares on $FB_{QS}>0$ (solar FIP biased, less active) stars are rare for the reasons described above, we look to expand our sample for figure~\ref{fg:FB_FQ_QS} to solar flares. A few solar flares were analyzed for abundances that included both high and low FIP elements: \citet{Feldman1990} analyzed a flare observed with Skylab and found significant abundance variations during the flare. In fact, the flare abundances were close to solar photospheric values. 
The elements used are O and Ne (high-FIP), and Mg and Ca (low-FIP). \citet{Schmelz1993} analyzed two flares observed by the Solar Maximum Mission (SMM) and \citet{Fludra1995} measured the abundances in these two flares again using a different method. In these flares, the elements observed were Ne and O (high-FIP) and Fe, Ca, Si, and Mg (low-FIP). 
Since quiescent abundances from the active regions were not directly measured by these authors, we use $FB_{QS} = 0.45 \pm 0.15$ for the quiescent solar corona representing roughly the typical enrichment by a factor of 2--4 of low-FIP elements in solar active regions \citep{Feldman1992}.
It can be seen in figure~\ref{fg:FB_FQ_QS} that the solar flares continue the trend of $FB_{FQ}>0$ shifting to $FB_{FQ}<0$ as $FB_{QS}$ increases. In other words, low-FIP enriched coronae such as the Sun, feature a relative enrichment of high-FIP elements during flares. 

\begin{figure}
\includegraphics[width=0.48\textwidth]{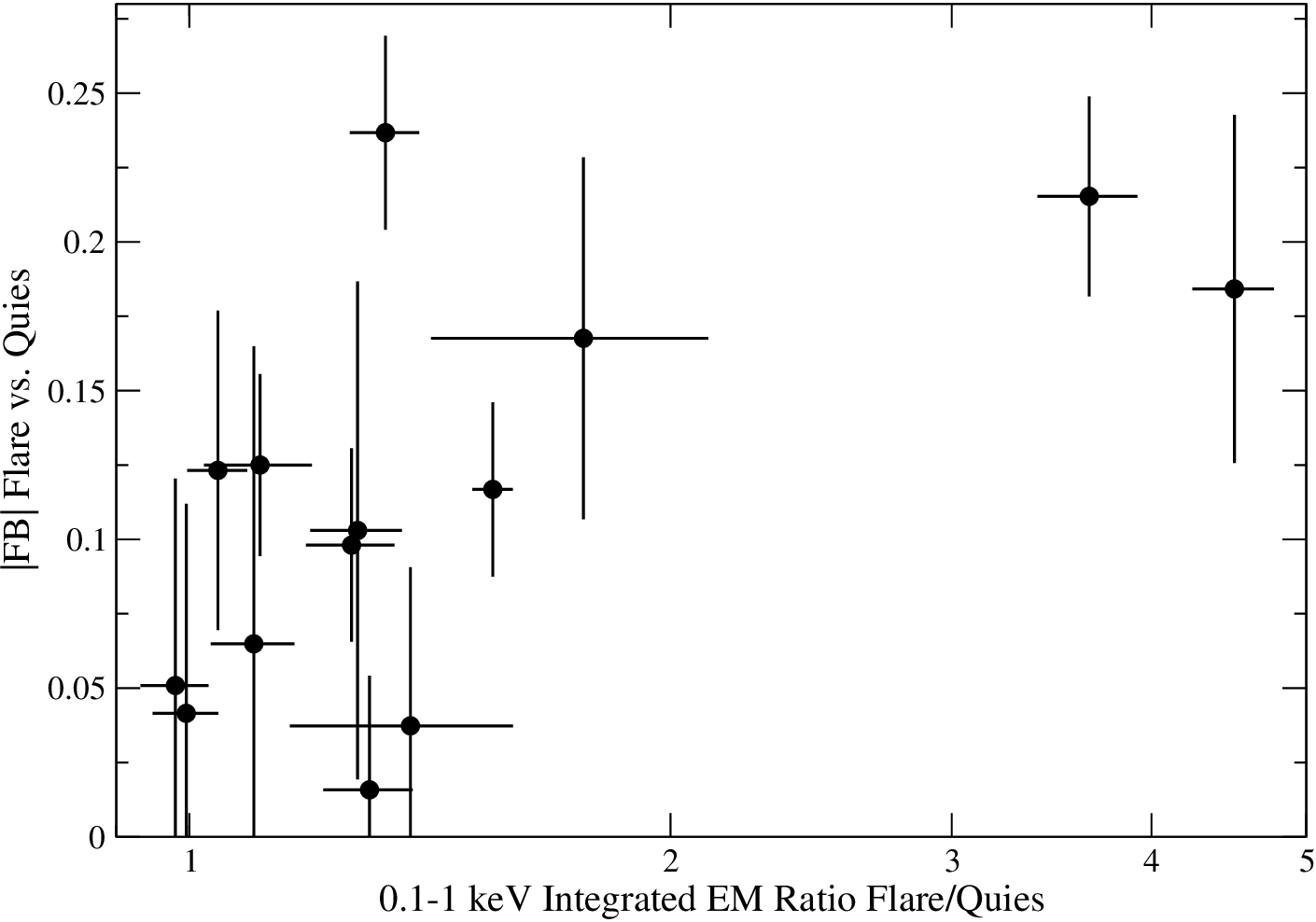}
\hspace{-20pt}
\caption{Absolute value of FIP bias measure $FB$ as a function of the ratio of integrated $EM$ from 0.1~to~1~keV in the flare relative to quiescence. A low-$T EM$ excess is required in order to enable the detection of abundance variations of the key elements O, Ne, Mg and Si.}
\label{fg:FB_1keV_EM}
\end{figure}

Why is the abundance variation not observed in all flares? That there is none, is certainly a possibility, but a better explanation is that our ability to detect variations is limited. Detection of abundance variation depends strongly on the excess of $EM$ during the flare, at temperatures that produce significant line emission. The emissivities of lines of O, Ne and Mg peak at temperatures below $kT~=$ 1~keV. Si ion emissivities peaks at slightly higher temperatures of kT$\sim$1.2~keV. At higher temperatures, these elements become fully ionized and emit no lines. Therefore, one needs a significant EM excess at temperatures of $\sim$1~keV to observe abundance variations. Looking at figure~\ref{fg:FeIEM}, we see that most of the EM excess during the flares appears at temperatures of $kT>$2~keV. Some of the flares show very little EM excess below 1~keV. For these, we expect difficulties in detecting abundance variations, 
given given the seemingly steady quiescence (background) emission.
Figure~\ref{fg:FB_1keV_EM} demonstrates this effect. It shows the $|FB_{FQ}|$ value vs. the flare to quiescence ratio of the FeIEM at 1~keV, which represents the EM excess during the flare from the relevant low-$T$ ($kT <$~1~keV) plasma. The figure shows very clearly that all of the flares in which no abundance variation was detected during the flare (FB$_{FQ}$ consistent with zero) "suffer" from little (to none) low-T, EM excess. When the EM excess is large, we always are able to detect variations. We conclude that the abundance change during the flare is set by the coronal and photospheric composition, but the actual measured value of $FB_{FQ}$ and our ability to detect the changes, depend on the EM excess at line emitting temperatures and thus on the details of the flare evolution. 

In some flares we therefore clearly detect a different composition than that of the quiescent corona (fig.~\ref{fg:FIP}). 
Chromospheric evaporation has been suggested as an explanation for varying abundances observed during flares, e.g. \citet{Ottmann1996}. In this scenario, accelerated fast particles from the reconnection region, spiral down the magnetic loop and are stopped in the denser chromosphere, releasing their energy. This heats chromospheric material to coronal temperatures and causes expansion, observed on the sun in the form of blue shifted lines during the initial phase of the flare \citep{Milligan2006}. During the decay phase, evaporation likely driven by heat conduction has been reported \citep{Schmieder1987, Zarro1988, Berlicki2005}. 
If indeed chromospheric evaporation is responsible for the excess emission of the flare, then the composition of the flaring plasma is expected to resemble photospheric composition. For a corona with a quiescent IFIP bias, evaporated chromospheric plasma would appear to be FIP biased with respect to quiescence, while for a FIP biased corona, the evaporated chromospheric plasma would produce an IFIP effect. Indeed, this is what is observed in figure~\ref{fg:FB_FQ_QS}.

In two of the flares, namely those on $\sigma$~Gem \citep{Guedel2002a} and on Proxima Cen \citep{Guedel2002b}, a Neupert effect \citep{Neupert1968}, has been observed. 
The Neupert effect is perceived as a direct evidence for chromospheric evaporation. The flare on $\sigma$~Gem, one of the most active RSCVn's with a strong quiescence IFIP bias $FB_{QS}=-0.55 \pm 0.08$, shows a clear {\it relative} enhancement of low-FIP abundances, reflected in a flare to quiescence FIP bias $FB_{FQ}=0.24 \pm 0.04$, \citep[see also][]{Nordon2006}. Conversely, Proxima Cen, a close by X-ray faint M dwarf with a small, if any, quiescent FIP bias of $FB_{QS}=-0.10 \pm 0.09$ shows the most distinct IFIP effect during the flare with $FB_{FQ}=-0.18 \pm 0.06$. The over all picture emerging from fig.~\ref{fg:FB_FQ_QS} is that flares bring up chromospheric material, which manifests the photospheric composition. If the coronal composition is different from that of the photosphere, the flare can produce a notable abundance effect, best observed by means of the X-ray emission lines.
Moreover, this result supports the viability of the quiescent IFIP effect, since it is those stars that also show a relative FIP bias during the flare. If the excess flare emission in these stars is due to evaporated chromospheric plasma and its composition reflects that of their chromosphere, it proves that indeed the coronal composition is different from the photosphere composition and that the inverse-FIP effect in these coronae is indeed genuine (and not just an artifact of using inappropriate photospheric abundances).

It still remains unclear how evaporation affects the observed abundances in the case of a micro-flaring quiescence. If the quiescence state is in itself a superposition of small flares, continuously evaporating material with photospheric composition, it should drive the mean coronal abundances toward photospheric. 
In order to accommodate the mean quiescent FIP bias with FIP variation during large flares, 
it requires that chemical mixing during micro-flares is insignificant, unlike the effect observed here in some of the large flares. Possible reasons for this include:
less energetic electrons not penetrating deep enough into the chromosphere and heating material of coronal composition, or less efficient heat conduction at the foot print, reducing conductive induced evaporation. There is also the possibility that typical micro-flares are fundamentally different from the large flares. On the sun we observe a large variety of flares in their geometries and behaviour (e.g. single loop vs. two ribbon, gradual vs. impulsive). It is therefore quite possible that the common micro-flare is not simply a scaled down version of the very large flares discussed in this work.

\subsection{Local abundances}
Variation in observed abundances during stellar flares does not necessarily require evaporation of plasma with different chemical composition. The solar corona is not uniform and abundances vary from one coronal structure to another. The FIP bias is also known to develop over a time scale of days, reaching the typical coronal values within 2--3 days and further increasing thereafter \citep{Widing2001}. Assuming stellar coronae are similar, the measured quiescent abundances are an EM-weighted average of all coronal structures. In a flaring state, the flare dominates the emission and the measured mean abundance will tend toward that of the flaring structure.

In order to explain figure~\ref{fg:FB_FQ_QS}, with similar solar-like coronal structures that develop a solar-like FIP effect with time, the flaring structures in the active coronae (those with lowest $FB_{QS}$ values) would have to be older than average, while in the less active coronae (highest $FB_{QS}$), they would have to be younger than average. However, It is not clear why this would be the case. On the other hand, the apparent correlation in figure~\ref{fg:FB_FQ_QS} between quiescent FIP bias and flare FIP bias, with no clear-cut counter examples of low-$FB_{FQ}$--low-$FB_{QS}$ or high-$FB_{FQ}$--high-$FB_{QS}$ data points, makes us deem the local abundance explanation fairly unlikely. Nonetheless, magnetic structure variations accompanied by a FIP/IFIP effect that changes with time, may have some effect on the measured $FB_{FQ}$ values causing it to vary from flare to flare (in addition to the other factors).

\section{Conclusions}
Eight large stellar flares, observed with \xmm, were analyzed for their thermal and chemical structure. With the six large flares observed with \cxc\ and analyzed using similar methods (paper 1), we present a sample of fourteen large flares with good photon statistics, observed in the X-ray at high spectral resolution. In order to identify and quantify abundance variations during the flares, that may be FIP related, we defined a FIP bias measure (FB).

In seven out of the fourteen cases, we find an enrichment of low-FIP elements compared to high-FIP elements during the flares (i.e., a solar-type FIP bias $FB_{FQ}>0$), five cases are consistent with no FIP-related bias and two show a preferential enrichment of high-FIP elements over the low-FIP ones (i.e, IFIP, $FB_{FQ}<0$). Note that since the abundance variation that we measure is between flaring and quiescent states of the coronae in the X-rays, no knowledge of photospheric abundances is required to determine $FB_{FQ}$. The FIP effect during the flare (the sign of FB$_{FQ}$) seems to correlate with the quiescent FIP bias of the corona, so that strong quiescent IFIP coronae tend to have $FB_{FQ}>0$, while coronae with lower values of $FB_{FQ}$ show the opposite effect. Solar flares also support this trend.

The most likely explanation for this effect is that of chromospheric evaporation during flares, where a significant amount of the excess EM during the flare is due to heated chromospheric material. For a FIP biased corona, chromospheric composition is relatively IFIP biased and vice versa: For IFIP quiescent coronae, chromospheric composition is relatively FIP biased.  This effect suggests that the observed inverse-FIP effect in the more active stellar coronae is indeed real and not an effect caused by photospheric abundances different from solar. Also, as flares seem to simply raise chromospheric material into the corona, the flares themselves can not be the direct cause for the IFIP effect - at least not the large flares observed here - as continuous strong flaring would tend to lower the absolute FB towards zero, but could not reverse the effect from FIP to IFIP or vice versa.

\begin{acknowledgements} 
This research was supported by ISF grant 28/03 and by a grant from the Asher Space Institute at the Technion.
\end{acknowledgements}


\Online
\begin{appendix}
\section{Line fluxes}
Fluxes of H-like ion lines include both transitions of the unresolved Ly-$\alpha$ doublet.
Fluxes of He-like ion lines include only the resonant transition.
Fluxes of L-shell Fe ion lines include all transitions within $\pm$0.03~\AA\ of the specified wavelength.
\begin{table}[!ht]
\caption{\label{tab:CC_Eri_lf} Measured line fluxes in 10$^{-4}$ photons s$^{-1}$ cm$^{-2}$ of CC Eri} 
\begin{tabular}{| lc | cccc |}
\hline
    &      &\multicolumn{2}{|c}{Quiescence} & \multicolumn{2}{c|}{Flare} \tabularnewline
Ion & Wave [$\AA$]& Flux   & Err & Flux & Err  \\
\hline
Fe XVII	&15.01	&2.49	&0.09	&3.26	&0.09\\
Fe XVIII	&14.21	&0.86	&0.06	&1.55	&0.05\\
Fe IX	&13.51	&0.57	&0.05	&1.04	&0.10\\
Fe X	&12.84	&0.51	&0.04	&0.84	&0.08\\
Fe XI	&12.28	&0.41	&0.03	&0.75	&0.11\\
Fe XII	&11.77	&0.26	&0.02	&0.44	&0.09\\
Fe XIII	&11.00	&0.23	&0.02	&0.24	&0.10\\
Fe XIV	&10.64	&0.22	&0.02	&0.16	&0.07\\
Fe XV	&1.85	&0.03	&0.03	&0.03	&0.01\\
Fe XVI	&1.78	&0.01	&0.05	&0.00	&0.01\\
C VI	&33.73	&-	&-	&5.92	&0.20\\
N VI	&28.78	&-	&-	&0.57	&0.06\\
N VII	&24.78	&-	&-	&3.63	&0.14\\
O VII	&21.60	&3.66	&0.41	&4.21	&0.14\\
O VIII	&18.97	&14.75	&0.51	&14.75	&0.20\\
Ne IX	&13.45	&2.96	&0.08	&3.70	&0.16\\
Ne X	&12.13	&4.96	&0.10	&6.59	&0.21\\
Mg XI	&9.17	&0.27	&0.01	&0.44	&0.05\\
Mg XII	&8.42	&0.28	&0.02	&0.59	&0.07\\
Si XIII	&6.65	&0.37	&0.02	&0.59	&0.04\\
Si XIV	&6.18	&0.28	&0.02	&0.48	&0.03\\
S XV	&5.04	&0.18	&0.02	&0.24	&0.02\\
S XVI	&4.73	&0.10	&0.02	&0.12	&0.02\\
Ar XVII	&3.95	&0.04	&0.01	&0.02	&0.01\\
Ar XVIII	&3.73	&0.02	&0.01	&0	&0.01\\
Ca XIX	&3.18	&0	&0.01	&0	&0.01\\
Ca XX	&3.02	&0.01	&0.01	&0	&0.01\\
Ni XIX	&12.43	&0.16	&0.02	&0.25	&0.05\\
\hline 
\end{tabular}
\end{table}

\begin{table}[!ht]
\caption{\label{tab:Algol_lf} Measured line fluxes in 10$^{-4}$ photons s$^{-1}$ cm$^{-2}$ of Algol} 
\begin{tabular}{| lc | cccc |}
\hline
    &      &\multicolumn{2}{|c}{Quiescence} & \multicolumn{2}{c|}{Flare} \tabularnewline
Ion & Wave [$\AA$]& Flux   & Err & Flux & Err  \\
\hline
Fe XVII	&15.01	&4.65	&0.13	&4.09	&0.30\\
Fe XVIII	&14.21	&2.83	&0.09	&2.81	&0.18\\
Fe IX	&13.51	&2.55	&0.18	&2.36	&0.35\\
Fe X	&12.84	&2.61	&0.17	&2.40	&0.33\\
Fe XI	&12.28	&2.31	&0.22	&2.32	&0.43\\
Fe XII	&11.77	&2.15	&0.22	&2.24	&0.41\\
Fe XIII	&11.00	&1.29	&0.26	&0.89	&0.47\\
Fe XIV	&10.64	&1.78	&0.20	&2.15	&0.43\\
Fe XV	&1.85	&0.32	&0.02	&0.90	&0.07\\
Fe XVI	&1.78	&0.00	&0.03	&0.25	&0.09\\
C VI	&33.73	&0.19	&0.15	&0.00	&0.30\\
N VI	&28.78	&0.58	&0.11	&0.67	&0.22\\
N VII	&24.78	&5.03	&0.20	&5.73	&0.41\\
O VII	&21.60	&1.21	&0.15	&1.26	&0.31\\
O VIII	&18.97	&9.63	&0.23	&10.32	&0.45\\
Ne IX	&13.45	&1.36	&0.24	&2.11	&0.47\\
Ne X	&12.13	&8.07	&0.35	&8.56	&0.67\\
Mg XI	&9.17	&0.92	&0.11	&0.96	&0.23\\
Mg XII	&8.42	&1.77	&0.17	&1.86	&0.36\\
Si XIII	&6.65	&1.08	&0.15	&1.03	&0.32\\
Si XIV	&6.18	&1.10	&0.26	&0.41	&0.56\\
S XV	&5.04	&0.55	&0.06	&0.20	&0.14\\
S XVI	&4.73	&0.35	&0.08	&0.48	&0.19\\
Ar XVII	&3.95	&0.24	&0.04	&0.00	&0.10\\
Ar XVIII	&3.73	&0.19	&0.05	&0.18	&0.14\\
Ca XIX	&3.18	&0.07	&0.03	&0.13	&0.08\\
Ca XX	&3.02	&0.01	&0.04	&0.27	&0.11\\
Ni XIX	&12.43	&-	&-	&0.43	&0.22\\
\hline 
\end{tabular}
\end{table}

\begin{table}[!ht]
\caption{\label{tab:47_Cas_lf} Measured line fluxes in 10$^{-4}$ photons s$^{-1}$ cm$^{-2}$ of 47 Cas.} 
\begin{tabular}{| lc | cccc |}
\hline
    &      &\multicolumn{2}{|c}{Quiescence} & \multicolumn{2}{c|}{Flare} \tabularnewline
Ion & Wave [$\AA$]& Flux   & Err & Flux & Err  \\
\hline
Fe XVII	&15.01	&2.36	&0.08	&2.84	&0.28\\
Fe XVIII	&14.21	&1.26	&0.05	&1.54	&0.16\\
Fe IX	&13.51	&1.02	&0.09	&1.65	&0.31\\
Fe X	&12.84	&0.79	&0.08	&0.65	&0.28\\
Fe XI	&12.28	&0.70	&0.10	&0.95	&0.39\\
Fe XII	&11.77	&0.53	&0.09	&1.04	&0.42\\
Fe XIII	&11.00	&0.16	&0.10	&0.23	&0.48\\
Fe XIV	&10.64	&0.27	&0.08	&0.98	&0.36\\
Fe XV	&1.85	&0.016	&0.005	&0.14	&0.05\\
Fe XVI	&1.78	&0.006	&0.009	&0.016	&0.075\\
C VI	&33.73	&1.24	&0.12	&1.10	&0.40\\
N VI	&28.78	&0.11	&0.05	&0.43	&0.21\\
N VII	&24.78	&0.78	&0.09	&1.22	&0.37\\
O VII	&21.60	&0.73	&0.08	&0.81	&0.38\\
O VIII	&18.97	&5.17	&0.14	&5.39	&0.42\\
Ne IX	&13.45	&0.99	&0.13	&0.02	&0.41\\
Ne X	&12.13	&2.70	&0.16	&3.74	&0.57\\
Mg XI	&9.17	&0.43	&0.05	&0.89	&0.20\\
Mg XII	&8.42	&0.61	&0.08	&0.75	&0.27\\
Si XIII	&6.65	&0.25	&0.03	&0.43	&0.10\\
Si XIV	&6.18	&0.20	&0.02	&0.23	&0.07\\
S XV	&5.04	&0.10	&0.01	&0.14	&0.05\\
S XVI	&4.73	&0.04	&0.01	&0.09	&0.06\\
Ar XVII	&3.95	&0.02	&0.01	&0	&0.04\\
Ar XVIII	&3.73	&0	&0.01	&0	&0.06\\
Ca XIX	&3.18	&0	&0.01	&0.03	&0.04\\
Ca XX	&3.02	&0	&0.01	&0	&0.06\\
Ni XIX	&12.43	&-	&-	&0.18	&0.18\\
\hline 
\end{tabular}
\end{table}

\begin{table}[!ht]
\caption{\label{tab:EK_Dra_lf} Measured line fluxes in
10$^{-4}$ photons s$^{-1}$ cm$^{-2}$ of EK Dra} 
\begin{tabular}{| lc | cccc |}
\hline
    &      &\multicolumn{2}{|c}{Quiescence} & \multicolumn{2}{c|}{Flare} \tabularnewline
Ion & Wave [$\AA$]& Flux   & Err & Flux & Err \\
\hline
Fe XVII	&15.01	&1.97	&0.05	&2.05	&0.18\\
Fe XVIII	&14.21	&0.92	&0.03	&1.14	&0.10\\
Fe IX	&13.51	&0.58	&0.06	&0.51	&0.19\\
Fe X	&12.84	&0.44	&0.05	&0.48	&0.15\\
Fe XI	&12.28	&0.49	&0.06	&0.34	&0.21\\
Fe XII	&11.77	&0.29	&0.05	&0.06	&0.21\\
Fe XIII	&11.00	&0.15	&0.06	&0.39	&0.26\\
Fe XIV	&10.64	&0.12	&0.05	&0.17	&0.18\\
Fe XV	&1.85	&0.019	&0.003	&0.07	&0.03\\
Fe XVI	&1.78	&0.01	&0.01	&0.03	&0.04\\
C VI	&33.73	&0.64	&0.07	&0.87	&0.23\\
N VI	&28.78	&0.07	&0.03	&0.06	&0.10\\
N VII	&24.78	&0.36	&0.05	&0.19	&0.14\\
O VII	&21.60	&0.52	&0.06	&0.52	&0.18\\
O VIII	&18.97	&2.34	&0.08	&2.65	&0.24\\
Ne IX	&13.45	&0.57	&0.08	&0.80	&0.27\\
Ne X	&12.13	&1.14	&0.10	&1.80	&0.34\\
Mg XI	&9.17	&0.24	&0.03	&0.34	&0.10\\
Mg XII	&8.42	&0.25	&0.04	&0.31	&0.14\\
Si XIII	&6.65	&0.15	&0.01	&0.31	&0.13\\
Si XIV	&6.18	&0.08	&0.01	&0.22	&0.29\\
S XV	&5.04	&0.04	&0.01	&0.08	&0.03\\
S XVI	&4.73	&0.01	&0.01	&0.02	&0.03\\
Ar XVII	&3.95	&0.007	&0.004	&0.04	&0.02\\
Ar XVIII	&3.73	&0	&0.01	&0.01	&0.03\\
Ca XIX	&3.18	&0.003	&0.003	&0.04	&0.02\\
Ca XX	&3.02	&0	&0.004	&0.03	&0.04\\
Ni XIX	&12.43	&0.13	&6.19	&0.35	&0.11\\
\hline 
\end{tabular}
\end{table}

\begin{table}[!ht]
\caption{\label{tab:HR1099_lf} Measured line fluxes in 10$^{-4}$ photons s$^{-1}$ cm$^{-2}$ of HR1099} 
\begin{tabular}{| lc | cccc |}
\hline
    &      &\multicolumn{2}{|c}{Quiescence} & \multicolumn{2}{c|}{Flare} \tabularnewline
Ion & Wave [$\AA$]& Flux   & Err & Flux & Err  \\
\hline
Fe XVII	&15.01	&5.13	&0.22	&5.35	&0.26\\
Fe XVIII	&14.21	&2.71	&0.15	&2.82	&0.15\\
Fe IX	&13.51	&2.45	&0.30	&2.14	&0.30\\
Fe X	&12.84	&2.23	&0.26	&2.26	&0.25\\
Fe XI	&12.28	&1.66	&0.34	&1.23	&0.35\\
Fe XII	&11.77	&1.11	&0.30	&1.40	&0.29\\
Fe XIII	&11.00	&0.62	&0.35	&0.56	&0.34\\
Fe XIV	&10.64	&0.43	&0.29	&0.80	&0.27\\
Fe XV	&1.85	&0.05	&0.02	&0.13	&0.02\\
Fe XVI	&1.78	&0.00	&0.03	&0.00	&0.02\\
C VI	&33.73	&4.57	&0.41	&5.46	&0.42\\
N VI	&28.78	&0.28	&0.17	&0.11	&0.16\\
N VII	&24.78	&3.33	&0.31	&2.86	&0.28\\
O VII	&21.60	&2.55	&0.31	&2.61	&0.29\\
O VIII	&18.97	&18.65	&0.48	&19.56	&0.46\\
Ne IX	&13.45	&4.03	&0.45	&4.98	&0.45\\
Ne X	&12.13	&16.34	&0.67	&17.07	&0.65\\
Mg XI	&9.17	&0.95	&0.17	&0.82	&0.17\\
Mg XII	&8.42	&1.46	&0.26	&1.86	&0.27\\
Si XIII	&6.65	&0.80	&0.23	&0.97	&0.24\\
Si XIV	&6.18	&0.48	&0.40	&1.02	&0.45\\
S XV	&5.04	&0.43	&0.05	&0.41	&0.05\\
S XVI	&4.73	&0.28	&0.07	&0.25	&0.06\\
Ar XVII	&3.95	&0.05	&0.03	&0.19	&0.04\\
Ar XVIII	&3.73	&0	&0.04	&0	&0.04\\
Ca XIX	&3.18	&0	&0.02	&0.05	&0.02\\
Ca XX	&3.02	&0	&0.03	&0	&0.03\\
Ni XIX	&12.43	&-	&-	&1.02	&0.17\\
\hline 
\end{tabular}
\end{table}

\begin{table}[!ht]
\caption{\label{tab:Prox_Cen_lf} Measured line fluxes in 10$^{-4}$ photons s$^{-1}$ cm$^{-2}$ of Proxima~Cen} 
\begin{tabular}{| lc | cccc |}
\hline
    &      &\multicolumn{2}{|c}{Quiescence} & \multicolumn{2}{c|}{Flare} \tabularnewline
Ion & Wave [$\AA$]& Flux   & Err & Flux & Err  \\
\hline
Fe XVII	&15.01	&1.08	&0.04	&5.53	&0.17\\
Fe XVIII	&14.21	&0.36	&0.02	&2.23	&0.09\\
Fe IX	&13.51	&0.21	&0.04	&2.30	&0.17\\
Fe X	&12.84	&0.03	&0.03	&1.75	&0.15\\
Fe XI	&12.28	&0.06	&0.04	&1.70	&0.19\\
Fe XII	&11.77	&0.05	&0.04	&1.07	&0.16\\
Fe XIII	&11.00	&0.00	&0.04	&0.52	&0.19\\
Fe XIV	&10.64	&0.00	&0.04	&0.75	&0.15\\
Fe XV	&1.85	&0.004	&0.004	&0.04	&0.01\\
Fe XVI	&1.78	&0.01	&0.01	&0.00	&0.02\\
C VI	&33.73	&1.87	&0.13	&6.19	&0.33\\
N VI	&28.78	&0.23	&0.04	&0.30	&0.10\\
N VII	&24.78	&0.64	&0.06	&2.78	&0.19\\
O VII	&21.60	&1.49	&0.08	&4.05	&0.21\\
O VIII	&18.97	&3.01	&0.09	&17.03	&0.32\\
Ne IX	&13.45	&0.57	&0.06	&2.42	&0.24\\
Ne X	&12.13	&0.53	&0.07	&6.02	&0.32\\
Mg XI	&9.17	&0.16	&0.02	&0.86	&0.09\\
Mg XII	&8.42	&0.09	&0.03	&1.03	&0.13\\
Al XII	&7.76	&0.02	&0.02	&0.21	&0.12\\
Al XIII	&7.17	&0.02	&0.02	&0.34	&0.12\\
Si XIII	&6.65	&0.06	&0.01	&0.86	&0.05\\
Si XIV	&6.18	&0.01	&0.01	&0.68	&0.04\\
S XV	&5.04	&0.002	&0.004	&0.27	&0.02\\
S XVI	&4.73	&0	&0.01	&0.18	&0.03\\
Ar XVII	&3.95	&0	&0.004	&0.01	&0.01\\
Ar XVIII	&3.73	&0	&0.01	&0	&0.02\\
Ca XIX	&3.18	&0	&0.004	&0	&0.01\\
Ca XX	&3.02	&0	&0.01	&0	&0.01\\
Ni XIX	&12.43	&0.25	&3.97	&0.28	&0.10\\
\hline 
\end{tabular}
\end{table}

\begin{table}[!ht]
\caption{\label{tab:Sig_Gem_lf} Measured line fluxes in 10$^{-4}$ photons s$^{-1}$ cm$^{-2}$ of $\sigma$~Gem} 
\begin{tabular}{| lc | cccc |}
\hline
    &      &\multicolumn{2}{|c}{Quiescence} & \multicolumn{2}{c|}{Flare} \tabularnewline
Ion & Wave [$\AA$]& Flux   & Err & Flux & Err \\
\hline
Fe XVII	&15.01	&4.15	&0.11	&6.02	&0.17\\
Fe XVIII	&14.21	&2.81	&0.08	&3.94	&0.11\\
Fe IX	&13.51	&2.15	&0.22	&4.27	&0.21\\
Fe X	&12.84	&2.85	&0.08	&4.41	&0.20\\
Fe XI	&12.28	&2.65	&0.07	&3.51	&0.26\\
Fe XII	&11.77	&1.80	&0.04	&3.79	&0.25\\
Fe XIII	&11.00	&1.46	&0.03	&2.30	&0.30\\
Fe XIV	&10.64	&1.86	&0.04	&7.25	&0.28\\
Fe XV	&1.85	&0.41	&0.04	&2.38	&0.03\\
Fe XVI	&1.78	&0.03	&0.03	&0.67	&0.04\\
C VI	&33.73	&10.33	&16.92	&5.25	&0.23\\
N VI	&28.78	&0.88	&1.15	&0.75	&0.12\\
N VII	&24.78	&5.71	&0.45	&6.86	&0.21\\
O VII	&21.60	&2.26	&0.24	&1.57	&0.18\\
O VIII	&18.97	&19.31	&0.50	&22.00	&0.29\\
Ne IX	&13.45	&3.84	&0.09	&2.84	&0.28\\
Ne X	&12.13	&15.88	&0.16	&22.78	&0.43\\
Mg XI	&9.17	&1.02	&0.03	&1.92	&0.15\\
Mg XII	&8.42	&1.98	&0.04	&5.31	&0.26\\
Al XII	&7.76	&0.11	&0.02	&1.25	&0.27\\
Al XIII	&7.17	&0.21	&0.02	&2.55	&0.26\\
Si XIII	&6.65	&1.11	&0.03	&2.43	&0.12\\
Si XIV	&6.18	&1.53	&0.03	&4.28	&0.10\\
S XV	&5.04	&0.62	&0.04	&1.80	&0.07\\
S XVI	&4.73	&0.52	&0.04	&2.28	&0.09\\
Ar XVII	&3.95	&0.41	&0.03	&0.46	&0.05\\
Ar XVIII	&3.73	&0.29	&0.03	&0.72	&0.07\\
Ca XIX	&3.18	&0.18	&0.02	&0.46	&0.04\\
Ca XX	&3.02	&0.03	&0.02	&0.14	&0.05\\
Ni XIX	&12.43	&0.43	&0.04	&0.59	&0.13\\
\hline 
\end{tabular}
\end{table}

\begin{table}[!ht]
\caption{\label{tab:Xi_Boo_lf} Measured line fluxes in 10$^{-4}$ photons s$^{-1}$ cm$^{-2}$ of $\xi$~Boo.} 
\begin{tabular}{| lc | cccc |}
\hline
    &      &\multicolumn{2}{|c}{Quiescence} & \multicolumn{2}{c|}{Flare} \tabularnewline
Ion & Wave [$\AA$]& Flux   & Err & Flux & Err \\
\hline
Fe XVII	&15.01	&4.71	&0.09	&5.90	&0.15\\
Fe XVIII	&14.21	&1.60	&0.04	&1.98	&0.07\\
Fe IX	&13.51	&1.03	&0.07	&1.16	&0.13\\
Fe X	&12.84	&0.62	&0.05	&0.86	&0.10\\
Fe XI	&12.28	&0.59	&0.07	&0.87	&0.13\\
Fe XII	&11.77	&0.23	&0.05	&0.42	&0.10\\
Fe XIII	&11.00	&0.18	&0.06	&0.26	&0.11\\
Fe XIV	&10.64	&0.00	&0.04	&0.00	&0.09\\
Fe XV	&1.85	&0.00	&0.01	&0.01	&0.02\\
Fe XVI	&1.78	&0.00	&0.01	&0.00	&0.03\\
C VI	&33.73	&1.96	&0.11	&2.08	&0.18\\
N VI	&28.78	&0.20	&0.04	&0.23	&0.07\\
N VII	&24.78	&0.61	&0.06	&0.60	&0.11\\
O VII	&21.60	&2.69	&0.10	&2.74	&0.16\\
O VIII	&18.97	&6.50	&0.12	&7.02	&0.20\\
Ne IX	&13.45	&1.37	&0.11	&1.50	&0.19\\
Ne X	&12.13	&1.32	&0.10	&1.69	&0.19\\
Mg XI	&9.17	&0.49	&0.03	&0.48	&0.06\\
Mg XII	&8.42	&0.29	&0.04	&0.28	&0.07\\
Si XIII	&6.65	&0.25	&0.03	&0.28	&0.07\\
Si XIV	&6.18	&0.08	&0.04	&0.15	&0.09\\
S XV	&5.04	&0.05	&0.01	&0.13	&0.02\\
S XVI	&4.73	&0.02	&0.01	&0.06	&0.02\\
Ar XVII	&3.95	&0.01	&0.01	&0.01	&0.01\\
Ar XVIII	&3.73	&0.02	&0.01	&0.02	&0.01\\
Ca XIX	&3.18	&0.01	&0.005	&0.005	&0.01\\
Ca XX	&3.02	&0.01	&0.005	&0.001	&0.01\\
Ni XIX	&12.43	&0.02	&0.04	&0.005	&0.07\\
\hline 
\end{tabular}
\end{table}

\end{appendix}

\end{document}